\documentclass[A4,reprint,cha]{revtex4-1}
\usepackage{graphicx,setspace}
\draft
\usepackage{dcolumn}
\usepackage{bm}

\begin{document}

\title{Canard-induced mixed mode oscillations in an excitable glow discharge plasmas}
\author{Md. Nurujjaman} \email{jaman\_nonlinear@yahoo.co.in}
\affiliation{Department of Physics, National Institute of Technology
  Sikkim, Ravangla, Sikkim-737139, India.} 
  \author{A. N. Sekar Iyengar}
\email{ansekar.iyengar@saha.ac.in} \affiliation{Plasma Physics
  Division, Saha Institute of Nuclear Physics, 1/AF, Bidhannagar,
  Kolkata -700064, India.}

\begin{abstract}
We demonstrated experimentally canard induced mixed mode oscillations
(MMO) in an excitable glow discharge plasma, and the results are
validated through numerical solution of the FitzHugh Nagumo (FHN)
model. When glow discharge plasma is perturbed by applying a magnetic
field, it shows mixed mode oscillatory activity, i.e., quasiperiodic
small oscillations interposed with large bounded limit cycles
oscillations. The initial quasiperiodic oscillations were observed to
change into large amplitude limit cycle oscillations with magnetic
field, and the number of these oscillation increases with increase in
the magnetic field. Fourier analysis of both numerical and
experimental results show that the origin of these oscillations are
canard-induced phenomena, which occurs near the threshold of the
control parameter. Further, the phase space plots also confirm that
the oscillations are basically canard-induced MMOs.
\end{abstract}

\maketitle
\section{Introduction}
\label{section:introduction}
It is well known that the change in a control parameters or external
perturbation in a threshold or an excitable systems produces various
nonlinear phenomena such as noise-induced resonances, canard
oscillations and mixed mode oscillation, which have been observed
experimentally as well as numerically in many physical, chemical,
biological and electronics systems
~\cite{PRL:Marino2004,PRE:Marino2006,PRL:Kramer2008,PRL:Makarov2001,PRE:Volkov2006,PRL:Mikikian2008,PRE:Marino2013,PRE:Marino2011,JChemPhys:Rao2011,Nature:McDonnell2011,Nature:Wiesenfeld1995,EPL:Gingl1995,PRL:Takeo2009,Scholarpedia:Wechselberger2007,Chaos:Borowski2010,J_Chem_phys:petrov1992,PhysicaD:koper1995}.
However, these kind of phenomena have been observed in very a few
experiments in case of
plasmas~\cite{PRL:Lin1995,POP:Dinklage1999,PRE:Nurujjaman2010,PRE:Nurujjaman2009,PRE:Nurujjaman2008,PRE:Nurujjaman2010}.
This is mainly because, it is not easy to achieve excitability
condition in a plasma system. So far most of the nonlinear dynamical
experiments, which depends on the excitability of plasma were
performed mostly in the glow discharge plasma, and in these
experiments the excitability has been achieved through Hopf
bifurcation~\cite{PRL:Lin1995,POP:Dinklage1999} or homoclinic
bifurcation~\cite{PRE:Nurujjaman2008}. Discharge voltage or discharge
current was the control parameter (CP) for the above mentioned plasma
experiments.  When plasmas is perturbed externally at its excitable
state by using noise or a periodic signal or together, the system
shows stochastic resonance, frequency entertainments, period pulling
and other perturbation-enhanced nonlinear
phenomena~\cite{PRL:Lin1995,POP:Dinklage1999,PRE:Nurujjaman2010,PRE:Nurujjaman2009,PRE:Nurujjaman2008,PRE:Nurujjaman2010a},
and one important feature of these experiments was that the coarse
change in the CP was sufficient to generate these kind of
phenomena. It is observed that if the change in the CP near the
threshold is very small, then such excitable system may also generate
canard-enhanced phenomena. For example, the FitzHugh Nagumo (FHN)
model or real experiments generates canard, and various
canard-enhanced phenomena due to small change in CP near the threshold
that have already been studied in
detail~\cite{PRE:Xiumin2007,PRL:Marino2004,J_Chem_phys:petrov1992,
  PRE:Volkov2006,Chaos:desroches2008,Chaos:brons2008,J_comp_neuro:vo2010,Nonlinearity:berglund2012,Chaos:muratov2008,arXiv:Touboul2013}.

The canard phenomena in an excitable system means the generation of
small quasiperiodic oscillations that has been observed through
numerical simulation as well as in a few experiments for a small
charge in the CP near the threshold of
excitability~\cite{PRL:Marino2004,PRE:Volkov2006,Scholarpedia:Wechselberger2007,Chaos:Borowski2010}.
An important feature of a system, which shows
canard-induced phenomena is that minute change in the CP produces
large bounded limit cycles of different frequency from small
quasiperiodic oscillations~\cite{PRE:Volkov2006}. Though minute change
in the CP is easily attainable for the numerical experiments to get
canard explosion, it is difficult to achieve such small change in the
CP for a real experiment. As the change required in the CP is for the
most of the experiments is order of the noise level, noise amplitude
suppresses the desired charge in the CP.  Another serious problem that
may also overshadow minute change in the CP is the parametric drift of
the plasma that keep on shifting the threshold point continuously.  In
case of glow discharge
plasma~\cite{PRE:Nurujjaman2010,PRE:Nurujjaman2009,PRE:Nurujjaman2008,
  PRE:Nurujjaman2010a}, where discharge voltage or current acts as a
CP, canard induced phenomena have not yet been observed by changing
the discharge voltage or current. This may be due to the fact that
discharge voltage or current acts as a coarse CP, and hence minute
change in the CP may be overshadowed by the presence of the noise and
parametric drift in such systems. It is observed that if an excitable
glow discharge plasma whose CP (in this case it was discharge voltage)
is kept fixed near the threshold, and a magnetic field is applied to
it, then the desired change in the CP can be achieved to get
canard-induced phenomena. One such canard-induced phenomena is the
mixed mode oscillation
(MMO)~\cite{Chaos:Borowski2010,J_Chem_phys:petrov1992,PhysicaD:koper1995,Chaos:muratov2008,arXiv:Touboul2013}.
Various mechanisms are responsible for MMOs in deterministic systems
like the existence of a Shilnikovtype homoclinic orbit or subcritical
Hopf bifurcation~\cite{SIAM:Desroches2012,PhysicaD:Simpson2011}.  In
glow discharge plasmas such kind of canard-induced MMO may also appear
near the Homoclinic bifurcations~\cite{PRE:Nurujjaman2008} as in this
case same conditions can be achieved that are generated by the
Shilnikovtype homoclinic orbit or subcritical Hopf bifurcation.

In this paper, we report the appearance of the canard-induced MMOs in
a glow discharge plasmas when the system was perturbed through a
magnetic field at an excitable state keeping discharge voltage
fixed. We observed two kinds of oscillations: small quasiperiodic
oscillations just after the introduction of the magnetic field, and
large bounded limit cycles with increase in the magnetic field. These
signals have a well-defined shape that seems similar to the
canard-induced mixed mode oscillations (MMOs).We also explored main
characteristics, typical frequencies and evolution of
inter-oscillations interval to understand the dynamics of the
system. These results are validated through the numerical simulation
of the FHN
system~\cite{PRE:Nurujjaman2010,PRL:Marino2004,PRE:Volkov2006}.  Phase
space plot also shows that these oscillations are MMO.

Rest of the paper has been organized as follows: we have discussed the
experimental set up and autonomous plasma dynamics in
Section~\ref{section:Experimental setup}.  The results and discussion
of the experiment and simulation results has been presented in
Section~\ref{section:result}. Finally a conclusion has been drawn in
Section~\ref{section:conclusion}.

\section{Experimental setup and autonomous plasma dynamics}
\label{section:Experimental setup}
\begin{figure}[ht]
\includegraphics[width=8.5 cm]{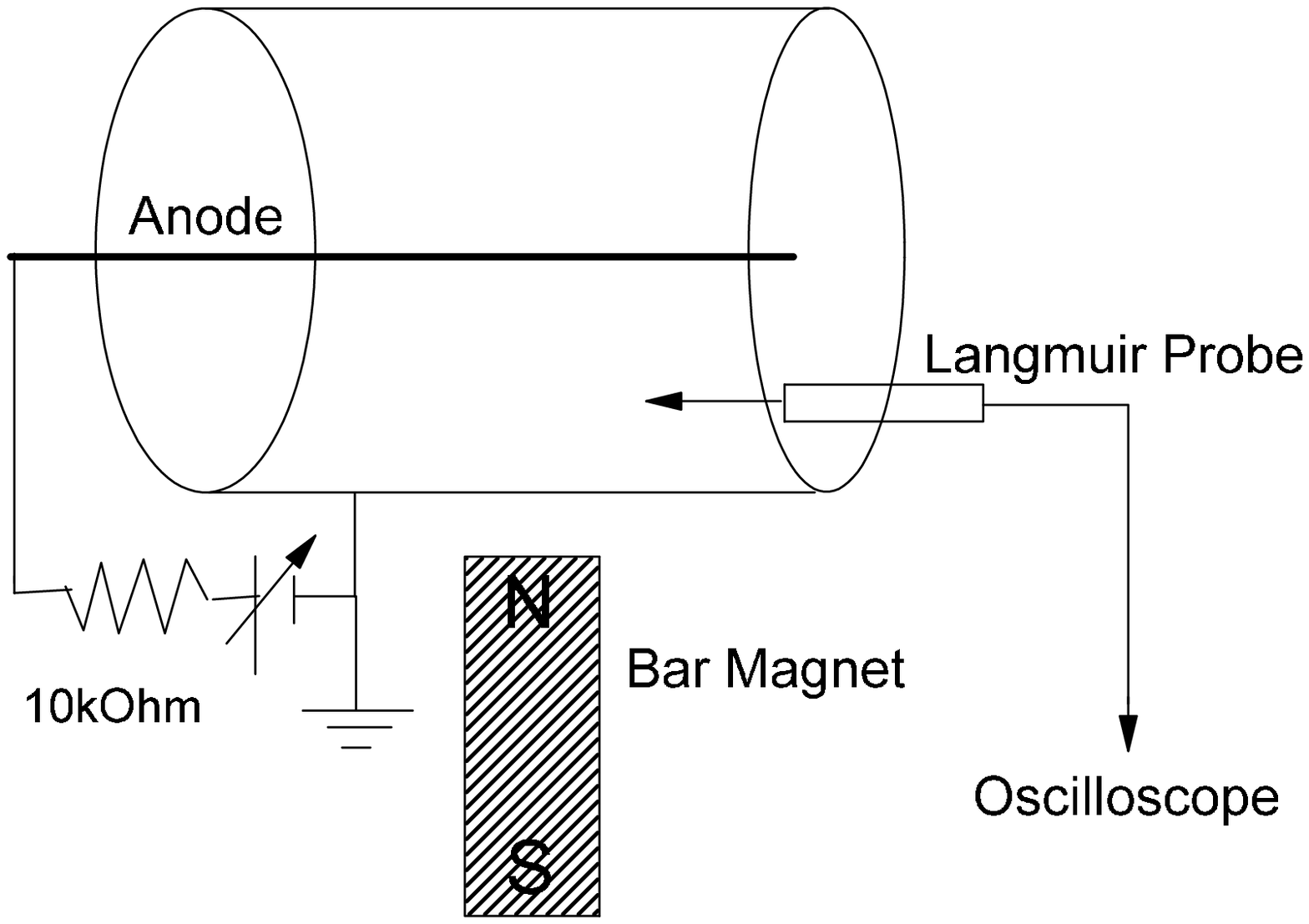}
\caption{Schematic diagram of the cylindrical
electrode system of the glow discharge plasma.
The probe was placed at a distance $l\approx12.5$ mm
from the anode.}
\label{fig1:setup}
\end{figure}

The experiments were performed in a hollow cathode dc glow discharge
plasma. The schematic diagram of the experimental setup is presented
in Fig~\ref{fig1:setup}. Here the experimental setup of
Ref~\cite{PRE:Nurujjaman2008} has been used. Detail of the
experimental condition will be found in Ref~\cite{PRE:Nurujjaman2008}.

The cylindrical hollow electrode [shown in Fig~\ref{fig1:setup}]inside
which plasma was generated, was kept inside a vacuum chamber and was
pumped down to a pressure around 0.001 mbar using a rotary pump. The
chamber was subsequently filled with argon gas at P= 0.36
mbar. Discharge was initiated by increasing the discharge voltage
(DV). At this pressure DV was 401 V to get the excitable dynamics.
The system observable was the electrostatic floating potential, which
was measured using a Langmuir probe used in
Ref~\cite{PRE:Nurujjaman2008}. Time series of the floating potential
has been recorded and analyzed to find out underlying dynamics. The
plasma density and the electron temperature were determined to be of
the order of 10$^7$cm$^{-3}$ and 3$-$4 eV respectively. Furthermore,
the electron plasma frequency was observed to be around 28 MHz,
whereas the ion plasma frequency was measured to be around 105 kHz. In
the present experiment, a magnetic field, which acted as control
parameter, was applied to the plasma by using a bar magnet as shown in
the same figure.

Through out the experiments, DV and pressure was kept constant. In all
the  experiments magnetic  field  was used  as  the control  parameter
(CP). In the excitable domain,  the system shows irregular and complex
oscillations at the initial stages  of the DV, and upon increasing the
DV, the  oscillations became regular  period-one oscillation.  Further
augmentation of the DV modified the oscillation profile and results in
the          induction          of         typical          relaxation
oscillations~\cite{PRE:Nurujjaman2008,Chaos:nurujjaman2007}, and these
oscillations  ceases  to  steady  state to  generate  excitable  state
through homoclinic  bifurcation that has  been discussed in  detail in
Ref~\cite{PRE:Nurujjaman2008}.  Once  the excitable state  is achieved
by changing the DV, it kept constant through out the experiment, and a
magnetic field is  applied to ensure small change  of the perturbation
to get canard-induced oscillations. From this point the magnetic field
acted as  CP, and the  magnetic field ensures  small change in  the CP
near the threshold to generate canard-enhanced phenomena.

\begin{figure}[ht]
\centering
\includegraphics[width=8.5cm]{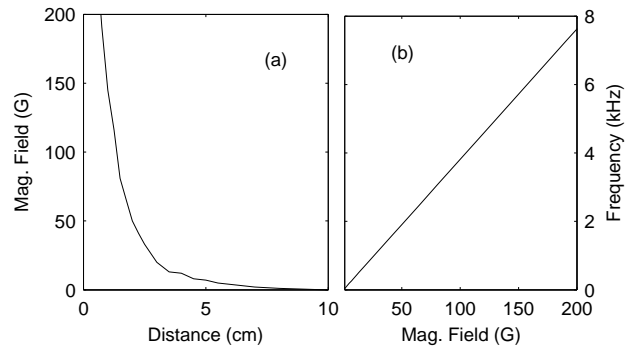}
\caption{Variation of the applied Magnetic field from the chamber.} 
\label{fig:mag}
\end{figure}
The variable of the external magnetic field, which acts as a CP
applied from the outside the chamber [Fig~\ref{fig1:setup}].  The
variation in the magnetic field with distance is shown in
Fig~\ref{fig:mag} (a). Fig~\ref{fig:mag} (b) shows the corresponding
ioncyclotron frequency ($f_{ci}$) of the plasmas that derived using
the relation $f_{ci}=1.52\times10^3Z\mu^{-1}B$ Hz, where,
$\mu=\frac{m_i}{m_p}$; Z is charge state and B is the magnetic field
in Gauss. $m_i$ is the argon mass and $m_p$ the mass of a proton. Once
the excitability is achieved through the change of the DV, it was kept
fixed through out the experiment and the magnetic field was varied to
get the desired dynamics. It is observed that the magnetic field
ensures the small change in the CP near the threshold to generate
canard explosion that was not possible by using DV. In the next
section, we have presented the experimental results with the applied
of magnetic field.

\section{Result and Discussion}
\label{section:result}

\begin{figure}[t]
\centering
\includegraphics[width=8.5 cm]{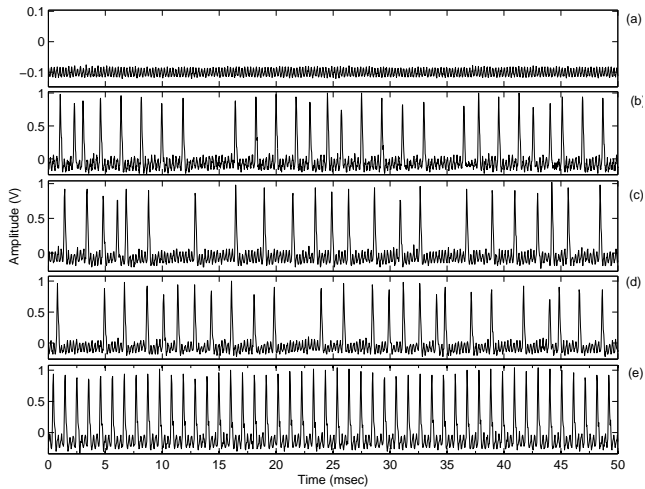}
\caption{Plasma oscillations at different magnetic fields. (a) small
 quasiperiodic oscillations at approximately $B=2$ Gauss. (b)-(d) Emergence of
 canard-induced oscillations with the application of magnetic field (4-20)  Guass
  Gauss), and (e) appearance of the regular mixed mode oscillation at B= 25 Gauss.}
\label{fig:plasma_oscil}
\end{figure}

For the experiments on canard-induced oscillations, the the DV was set
to 401 V for the gas pressure $P=0.36$~mbar so that the output of the
plasma floating potential showed excitable fixed point behavior in the absence
of magnetic field. The set point was kept a little away from the threshold so
that the system remains in a stable state under the influence of parametric
drifts and absence of the magnetic field. At this point magnetic field was
applied by using a bar-magnet and its intensity was varied through the
variation of spatial distance from the anode. Variation of the field with
spatial distance has already been shown in Fig~\ref{fig:mag}(a).

The DV was set at V= 401 V during the experiment, and at this point
plasma showed constant floating potential in the absence of the
magnetic field perturbation. Various kind of oscillations depending on
the minute change in the magnetic field near the threshold
point. Fig~\ref{fig:plasma_oscil} shows the plasma floating potential
oscillations for different magnetic
field. Fig~\ref{fig:plasma_oscil}(a) shows the quasiperiodic small
oscillations at a magnetic field (B) of $\approx 1-2$ G (i.e., just
after the introduction of the magnetic
field). Fig~\ref{fig:plasma_oscil}(b) shows the same plasma
fluctuations at $B=4~G$. It shows that the large but bounded periodic
limit cycle oscillations appears between the small quasiperiodic
oscillations. Usually there are 5 small oscillations between two large
oscillations. Sometimes, 4 or 3 small quasiperiodic oscillations were
also observed. There are also long sequence of small quasiperiodic
periodic oscillations in between two large sporadic periodic
oscillations. Appearance of sporadic long quasiperiodic sequence may
be due to parametric drifts of the system from the mean fixed
point. These oscillations were observed for wide range of magnetic
field (4 G to 20 G) that is shown in
Fig~\ref{fig:plasma_oscil}(b)$-$(d). When the magnetic field became 25
G, large oscillations were observed after every two small oscillations
[Fig~\ref{fig:plasma_oscil}(e)] and this has been observed till 100
Gauss. As the magnetic field applied from outside the vacuum vessel,
it was not possible going beyond the 100 G limit with the present
configuration of the experimental setup.

\begin{figure}[t]
\centering
\includegraphics[width=8.5cm, height=5 cm]{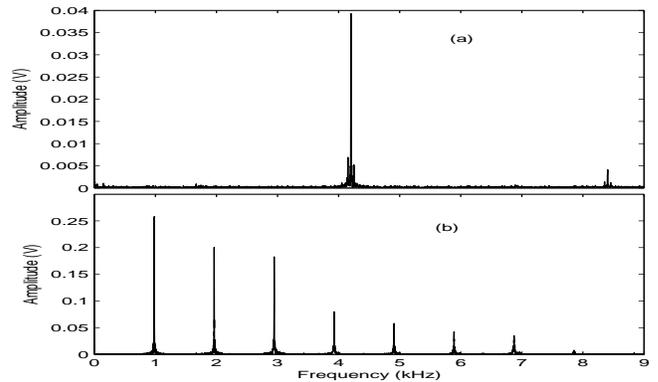}
\caption{FFT of experimental data: (a) shows the FFT of
 quasiperiodic plasma oscillations and (b)oscillations after appearance
 of canard phenomena. They shows the frequency of quasiperiodic
 plasma oscillations is approximately 4 times higher than canard oscillations.}
\label{fig:plasma_fft}
\end{figure}

 \begin{figure}[ht]
\centering
\includegraphics[width=8.5cm,height=5 cm]{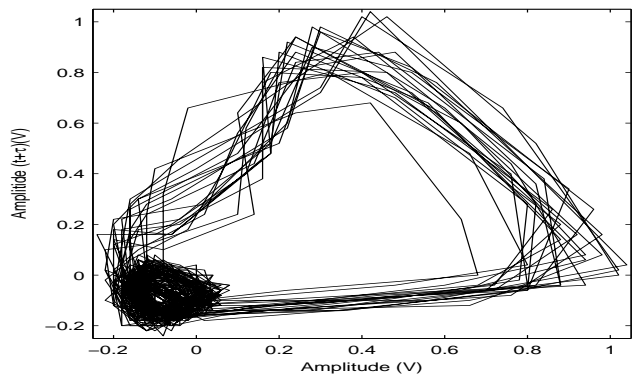}
\caption{Phase space plot of canard-induced oscillations obtained from
  experiment. It shows the emergence of large limit cycles oscillation in between
  small quasi-periodic oscillations. Smaller oscillations are contaminated
  by noise. Delay $\tau=0.04$ ms.}
\label{fig:plasma_phase}
\end{figure}

Fig~\ref{fig:plasma_fft} (a) and (b) show respectively the Fourier
transform of the large amplitude and quasiperiodic small amplitude
plasma potential oscillations. Frequency of the small quasiperiodic
oscillations and large amplitude limit cycle oscillations are around
4.2 kHz and 1.0 kHz respectively.  
\newpage
Frequency of the small
quasiperiodic oscillations is almost 4 times greater than the large
oscillations.  Fig~\ref{fig:plasma_fft}(a) shows that the frequency of
the quasiperiodic small oscillations is almost multiple and 4 times
higher than that of the large amplitude limit cycle oscillations and
this is consistent with the experimental observation of the
canard-induced
oscillations~\cite{PRL:Marino2004}. Fig~\ref{fig:plasma_phase} shows
the phase space plot of the experimental data, and it shows the
generation of the canard-enhanced trajectory. Moreover, the phase
space trajectory clearly shows the canard structure present in the
system. As there are are small quasiperiodic periodic oscillation
followed by a number of large amplitude limit cycle oscillation, these
phenomena are termed as mixed mode oscillation (MMO).

The main feature of a system which show MMO is that it must be
nonlinear with multiple
timescale~\cite{Chaos:desroches2008,Chaos:brons2008,Chaos:muratov2008}. Occurrence
of the multiscale dynamics in an excitable plasma system is already
been confirmed~\cite{PRE:Nurujjaman2008}, where multiscale dynamics
has already been exploited to demonstrate noise-induced coherence and
stochastic
resonances~\cite{PRE:Nurujjaman2008,PRE:Nurujjaman2010a}. As in the
case of noise- induced phenomena, the same system was observed to
undergo homoclinic bifurcation~\cite{PRE:Nurujjaman2008}, it may be
concluded that the origin of the MMO is canard-enhanced phenomena
which were generated due to homoclinic bifurcation after the
application of the magnetic field perturbation.

In the next section numerical simulation of FHN model is presented to
validate our experimental results.

\section{Numerical results: Model Analysis}
\label{section:model}
To validate our experimental results, numerically simulation has been
carried out of the FitzHugh Nagumo (FHN) model. Earlier same model has
also been used to validate the noise-induced resonances in case of
glow discharge plasma experiments~\cite{PRE:Nurujjaman2010}, where the
following FHN model~\cite{PRE:Volkov2006,PRE:Nurujjaman2010} has been
studied for an excitable system, whose equations of motion are
\begin{figure}[ht]
\centering
\includegraphics[width=8.5cm,height=5 cm]{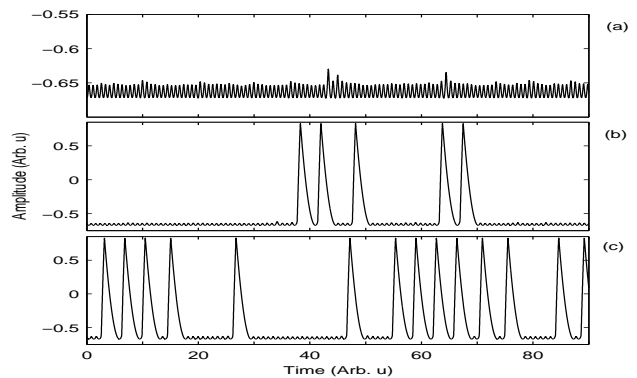}
\caption{Oscillation in FHN model. (a) small quasiperiodic
  oscillations at a1 = 0.99880 (b) Emergence of few canard-induced
  oscillations (large amplitude) a2 = 0.99950 and (c) a3 = 0.99999.}
\label{fig:simulation}
\end{figure}

\begin{figure}[ht]
\centering
\includegraphics[width=8.5cm,height=5 cm]{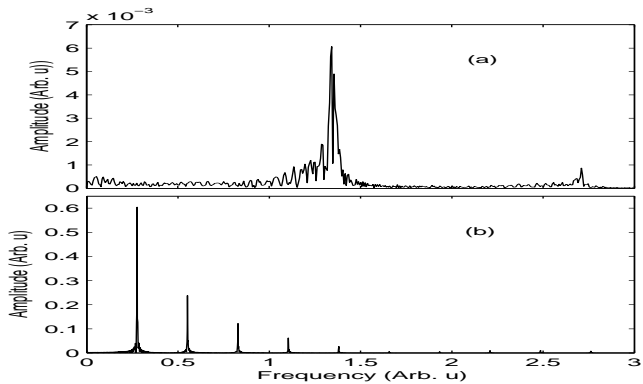}
\caption{FFT of the simulated data: (a) shows the FFT of quasiperiodic
  plasma oscillations and (b)oscillations after appearance of canard
  phenomena. They shows the frequency of quasiperiodic plasma
  oscillations is also approximately 4 times higher than small
  oscillations. $y$-axis of the Figs (c) is scaled down by 100.}
\label{fig:fft_sim}
\end{figure}

 \begin{eqnarray} \frac{dx}{dt}&=\frac{1}{\epsilon}(x-\frac{x^3}{3}-y)\nonumber\\ 
 \frac{dy}{dt}&=x+a \end{eqnarray}
 
 where, $\epsilon=0.01$ and the control parameter (`$a$') governs the
 dynamics. here $a$ is the control parameter, and it plays the same
 role of the magnetic field in the present experiment. For $|a|>1$ and
 $|a|<1$, the system shows fixed point and limit cycle oscillations
 respectively. Therefore, $a_{th}=1$ is the threshold and the system
 shows fixed point and oscillatory behavior above and below this point
 respectively. Frequency of the limit cycle oscillations increases
 between $0\leq a<1$ and then decreases between $0\geq a>-1$. Detail
 behavior with respect to an excitable plasma system has been
 discussed in Ref~\cite{PRE:Nurujjaman2010}.
 
When $a=0.99880$ the system shows quasiperiodic small oscillations as
shown in Fig~\ref{fig:simulation}(a), and the frequency of the
oscillations is shown in Fig~\ref{fig:fft_sim}(a). When $a$ is
increased by a small amount [$a=0.99880$], the system start showing
rapid but bounded growth in the limit cycle oscillation as shown in
Fig~\ref{fig:simulation}(b) and such growth is termed as canard
explosion. No. of limit cycles oscillations increases with increase in
the CP value that is typical for canard
oscillations~\cite{PRE:Volkov2006}.  Phase space plot also shows clear
the appearance of canard structures. As the amplitude of the model
solutions are large compared to the quasiperiodic small oscillations,
expanded version is shown in lower panel [Fig~\ref{fig:simulation}
  (b)]. From the time series and phase space plot it is clear that the
system show MMO.

 \begin{figure}[ht]
\centering
\includegraphics[width=8.5cm,height=5 cm]{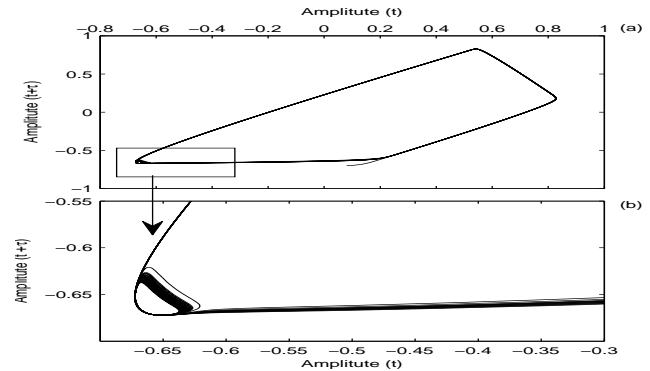}
\caption{Phase space plot of canard-induced oscillations obtained from
  simulation data. It shows the emergence of larger limit cycles from small
  quasi-periodic oscillations. Time delay $\tau=0.3$.}
\label{fig:plasma_phase,height=5 cm}
\end{figure}

\section{Conclusion}
\label{section:conclusion}
Effect of magnetic field near the threshold of an excitable plasma
system has been studied. Dynamics of the above system is multiscale in
nature, and this has already been exploited to get noise-induced
resonances using same
system~\cite{PRE:Nurujjaman2008,PRE:Nurujjaman2010}. When the same
system is perturbed by a magnetic field, it shows canard-induced mixed
mode oscillations.  Once again the multiscale dynamical behavior has
been exploited successfully by applying magnetic field.  The result
has also been validated through numerical simulation of the FHN model.

Beyond the interest of the study of these nonlinear phenomena from
experimental and dynamic point of view, their characterization is also
very important for experiments involving real application in glow
discharge plasma. Such study may be useful for various application of
discharge plasma. The theoretical analysis of such dynamics from
actual plasma dynamics may be the subject of future works.

\section*{Acknowledgement}
One of the authors (MN) appreciate the valuable comments and
suggestions of Martin Wechselberger on the experimental results. MN
acknowledges the constant support and encouragement from the Director,
NIT Sikkim. Both the authors also like to thank D. Das, S. S. Sil and
A. Bal of the Plasma Physics Division for their help during the
experiments. 
\bibliographystyle{aipnum4-1}
\bibliography{mybib}  

\end{document}